\documentstyle[11pt,epsfig]{article}
\def \vec #1{\mbox{\boldmath ${#1}$}}

\def \be{\begin{equation}}
\def \ee{\end{equation}}
\def \bea{\begin{eqnarray}}
\def \eea{\end{eqnarray}}
\def \ld{\left\langle}
\def \rd{\right\rangle}

\def \Jij{J_{ij}}

\def \Eh{{1\over2}}

\def \Nh{{N\over 2}}
\def \cH{{\cal H}}
\def \cO{{\cal O}}
\def \cP{{\cal P}}

\def \vS{\vec{S}}

\begin{document}
\title{\bf Optimization by Move--Class Deflation}
\author{Reimer K\"uhn${}^1$\thanks{Supported by a Heisenberg fellowship}, 
Yu-Cheng Lin${}^1$, and Gerhard P\"oppel${}^2$ \\{}\\
${}^1$ Institut f\"ur Theoretische Physik, Universit\"at Heidelberg\\
Philosophenweg 19, 69120 Heidelberg, Germany\\
${}^2$ Uhlandstrasse 8, 93049 Regensburg, Germany}

\date{May 10, 1998}

\maketitle

{\bf Abstract:} A new approach to combinatorial optimization based on
systematic move--class deflation is proposed. The algorithm combines
heuristics of genetic algorithms and simulated annealing, and is mainly
entropy--driven. It is tested on two problems known to be NP hard, namely 
the problem of finding ground states of the SK spin--glass and of the 
3-$D$ $\pm J$ spin--glass. The algorithm is sensitive to properties of 
phase spaces of complex systems other than those explored by simulated 
annealing, and it may therefore also be used as a diagnostic instrument.
Moreover, dynamic freezing transitions, which are well known to hamper 
the performance of simulated annealing in the large system limit are
not encountered by the present setup.

\section{Introduction}

Standard wisdom has it that strictly cost--decreasing algorithms are a bad
choice for solving `hard' optimization problems characterised by having
huge numbers of locally optimal yet globally suboptimal solutions. The 
travelling salesman problem or the problem of finding ground--states of 
disordered frustrated systems such as the SK and the 3-$D$ $\pm J$ 
spin--glasses are well known to belong to this category (for an overview, 
see \cite{GreSou}; another such problem, which has recently attracted some 
attention, is the binary--perceptron problem \cite{KrMe,Ho92,Pat93}). 
A common feature of these problems is that they have discrete phase spaces
whose volume grows exponentially (or faster) with problem size, so that a
complete enumeration of the universe of possible solutions as a means of
finding the optimum is generally unfeasible. In terms of computational 
complexity \cite{GaJo}, the three problems just mentioned are indeed 
known to be NP hard.

A number of algorithms have been invented to avoid getting trapped in local
minima of cost- or energy functions --- the most prominent and versatile
among them being perhaps the simulated annealing algorithm \cite{KiGe,Kir}. 
This approach introduces a mechanism of thermal activation accross barriers as
implemented in the Metropolis algorithm \cite{MRRTT} to escape from local 
minima of the cost function. The algorithm generates a Markov chain that 
approaches thermal equilibrium at a given temperature (if granted sufficient 
time to evolve), and as the temperature is gradually lowered to zero, thermal 
equilibrium singles out the ground state of the system, or one of the ground
states in case of degeneracies. A problem with this algorithm, though, is that
equilibration times at low temperatures may be excessively large in large
systems, and that the algorithm may {\em effectively}\/ get trapped in
suboptimal regions of configuration space because of time constraints. The
design of efficient cooling schedules such as to make this at least an
unlikely event has therefore always been of major concern of those working
in the field. As a rule of thumb, it may be said that to devise an efficient
cooling strategy always requires some tayloring depending on the system being
investigated. 

The so-called threshold--accepting algorithm \cite{DueSch} can
be thought of as a variant of simulated annealing, the main difference being
that unfavourable moves are accepted with uniform probability up to some
threshold which is then gradually decreased to zero. 

In population based algorithms of which the genetic ones \cite{GenAlg} are 
perhaps the best known, a somewhat different approach is taken. Here 
a number of pairs within a `population' of in general suboptimal solutions 
is selected and combined to form offspring which inherits usually equal 
parts of the solution of the optimization task from either parent. 
Specifically, if the solution  to a complex optimization task is encoded
in a bit-string of, say, length $N$, then the offspring will inherit 
$\cO(N/2)$ of the bits from one parent, and the remainig ones from the 
other, just as in sexual reproduction. The quality of the offspring is 
evaluated, and of the total new population now including the offspring, 
a certain fraction representing the fittest is retained to go for a new 
round of random pairing, offspring production and selection. Quite often,
the genetic crossover process just described is combined with a certain rate
of mutations, i.e. random changes of single bits in the copying process.
Changes which involve mutation without crossover are also sometimes 
considered.

For things to come is useful to characterise the algorithms just described 
by their move class. In terms of a {\em distance\/} in phase space, as
measured, e.g. by the Hamming distance of two bit-strings representing
different solutions or -- in spin-systems -- by the number of overturned 
spins in a single move, the simulated annealing or threshold-accepting
type algorithms are typically based on a set of {\em local\/} moves exploring
distances $d=\cO(1)$ in phase space by a single move, whereas in genetic
type algorithms the moves are non-local with $d=\cO(N/2)$, sometimes
mixed with $d=\cO(1)$ moves in cases using mutation without crossover.
 
In the present paper, we propose an alternative approach to complex 
optimization problems, based on a systematic tuning of move--classes from
{\em macroscopic}, though usually $o(N)$, to microscopic \cite{KuPoe}. 
The strategy will
be refered to as optimization by move--class deflation (OMCD) in what follows. It is perhaps best explained in terms of a specific example, viz. the 
problem of finding the ground state(s) of spin--glass Hamiltonians, such 
as that of the SK spin-glass \cite{SK} or that of the 3-$D$ $\pm J$ spin-glass
\cite{EA}. This will be done in  Sect. 2, where we explain the heuristics,
and verify it for the spin--glass problems, and analytically some aspects of 
it also on the toy problem of finding the ground-state of a Curie--Weiss 
ferromagnet. We sall see that OMCD makes {\em constructive use\/} of a
feature which is generally considered as one of the most difficult problems 
to be overcome, namely of the usually very high dimensionality of the phase 
spaces in typical combinatorial optimization problems. Specifically, the
feature we are able to {\em exploit\/} is  the fact that local densities of
state typically scale exponentially in system size.

In OMCD, the move--class deflation schedule plays a role analogous to the 
annealing schedule in the simulated annealing algorithm. Of particular 
interest here is the scaling of the initial size of the moves with the 
problem dimension $N$. This problem is dealt with in Sect. 3. In Sect. 4
we present results on ground--state energies of the spin--glass problems
introduced in Sect. 2, and on their scaling with system size. The MCD-algorithm
is sensitive to  properties of phase spaces of complex systems other than
those `seen' by simulated annealing, and it may thus be used as a new device
for phase space diagnostics. This aspect is described in Sect. 5.
Sect. 6 will conclude our paper with a summary and discussion.

\section{The Heuristics of OMCD}

Let us consider the problem of finding the ground state(s) of a spin--glass
Hamiltonian such as
\be
\cH_N = - \sum_{(i,j)} \Jij S_i S_j\ ,
\label{HN}
\ee
where the $S_i$, $1\le i \le N$, denote Ising--spins with values 
in $\{\pm 1\}$. This Hamiltonian may represent an SK spin--glass \cite{SK} 
in which case the sum in (\ref{HN}) extends over all $N(N-1)/2$ pairs of 
the system and the $\Jij$ are assumed to be Gaussian random variables of mean 
$\ld \Jij\rd = J_0/N$ and variance $\langle \Jij^2 \rangle - 
\ld \Jij\rd^2 =1/N$. Alternatively, one may consider a model with 
short--range interactions such as the 3-$D$ $\pm J$ spin--glass \cite{EA}, 
in which case the sum in (\ref{HN}) extends over all nearest neighbour 
pairs of a 3-$D$ simple cubic lattice, and the $\Jij$ randomly take values 
in $\{\pm 1\}$.

Our strategy to find ground--states of (\ref{HN}) is as follows. We start 
from some randomly chosen initial spin configuration $\vS_0$. Then we select a 
randomly chosen subset of the spins, containing (on average) $d$ spins.
Typically, $1 \ll d \ll N$. An attempt is made to flip these spins {\it 
simultaneously}. The attempt is accepted {\it only}, if it does not lead to 
an increase of $\cH_N$, otherwise it is rejected. After this procedure has 
been repeated many times, the move--class is systematically `deflated', 
e.g., by reducing the (average) size of the sets of spins suggested for a 
simultaneous flip by 1 ($d \to d - 1$), or by a certain factor ($d \to \gamma 
d$, with $\gamma < 1$). And the procedure of proposing and accepting/rejecting 
simultaneous spin--flips depending on whether $\cH_N$ decreases or not 
continues with this smaller move class. Eventually the move--class is 
reduced to consist of single spin flips only, corresponding to 
zero--temperature Monte--Carlo dynamics. The scheme is characterised by
the number of attempts at any given size of move--class and the way in 
which move--classes are deflated; this constitutes the move--class deflation 
schedule, which plays a role analogous to the cooling schedule in simulated annealing.

Why do we expect this scheme to give us good candidates for low--energy
states and, given enough computer time, even ground--states? First, it is 
clear that by allowing macroscoping moves, we can jump across energy barriers.
If we allow occasional big moves to the very end, still allowing only moves
that decrease (1), we are even {\it guaranteed}\/ to find the bottom of the 
deepest energy valley eventually. 

But why can we expect our method to be efficient? Well, this is a delicate 
question, and, as a matter of fact, we cannot expect this under all 
circumstances, e.g., when the energy landscape has a golf--course topology: 
almost everywhere flat with occasional small and narrow dips (optimization 
is generally difficult and slow in such situations). But for energy landscapes 
which are {\em normal\/} in the sense that, roughly speaking, deep valleys 
are also wide valleys, the following argument shows that our strategy for
finding low--energy configurations should be an efficient one, even if the
energy surfaces are rough in the sense of exhibiting  complicated ``valleys
within valleys within valleys..." structures, which appears to be quite
common for NP hard problems.

The first observation to make is that configuration spaces of the systems
we are considering are usually characterised by having very high dimension.
This circumstance, among others, is precisely what makes optimization 
difficult for these systems. For us {\em it is the staring point of our
strategy}. One of the main consequences of high dimensionality is that 
almost all states of these systems have energies which are neither very high 
nor very low. The reason is that states of {\em typical\/} energy are near
the ``surface" of those regions in configuration space which have either rather
high or rather low energy, and that virtually all volume of high dimensional 
objects is well known to be concentrated near their surface. In physics
terms, states of typical energy have the largest (local) density of states
(DOS). Since local densities of state typically scale exponentially in system size, states having an energy that differs significantly, i.e. on an extensive
scale, from typical are exponentially (in system size $N$) less probable 
than those having typical (hence average) energy. This observation is clearly
confirmed in practice, and it might be called the {\em message of 
initialization}: picking an initial state $\vS_0$ of (\ref{HN}) at random, i.e.
uncorrelated with the $J_{ij}$, we will {\em almost surely\/} find it to
have zero energy (per spin) in the large system limit.

Running the OMCD algorithm, we start by making random macroscopic (or perhaps
rather mesoscopic) moves which are accepted if they do not increase the
energy. Were do we get? At states which have both, a high probability --- 
otherwise they would not have been selected at random --- and a lower energy,
because only moves which lower (or at least do not increase) the energy are accepted. Such states may be expected to be near the surface (``half up") of both wide and deep valleys. Eventually, the macroscopic moves suggested will
no longer be accepted, because being macroscopic, they would lead us outside 
or higher up the wide and deep valley we have already found. The probablility 
of selecting some state which may be deep inside some other (narrower and in 
the end suboptimal) valley --- that is, lower in energy than half up the deep 
and wide valley we have already found --- can be considered negligibly small 
in the large system limit because of entropy (density of state) considerations 
as discussed above. Next, the move--class is reduced, so that states {\em 
within\/} the valley we have already selected can be reached and are accepted 
if they lower (or do not increase) the energy. By the same argument as before, 
we may expect to arrive at states near the surface of the deepest and widest 
valley structures {\em inside\/} the valley we have already selected. 
Moreover, since the move--class contains only smaller moves, a larger --- 
and by our reasoning irrelevant --- part of the phase space is already 
effectively shielded from our search. Having arrived at this finer level, 
we can repeat our argument as before, with a move class that is reduced once 
more. By systematically reducing the move class, we thus explore the energy 
landscape down to the deepest and finest structures, shielding off larger and 
larger parts of phase space from our search as we go along.

\begin{figure}[t]
{\centering 
\epsfig{file=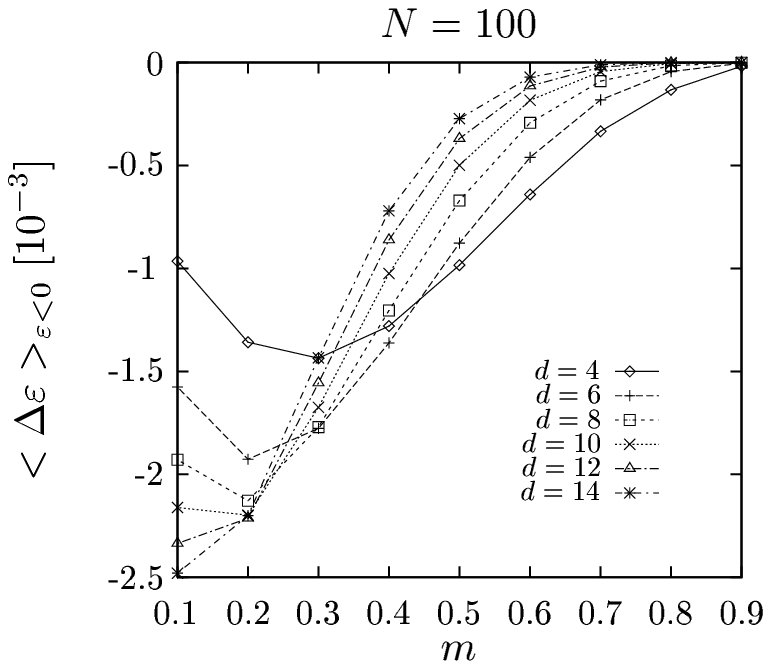, width=0.475\textwidth, height=5cm}  
\hfill{}
\epsfig{file=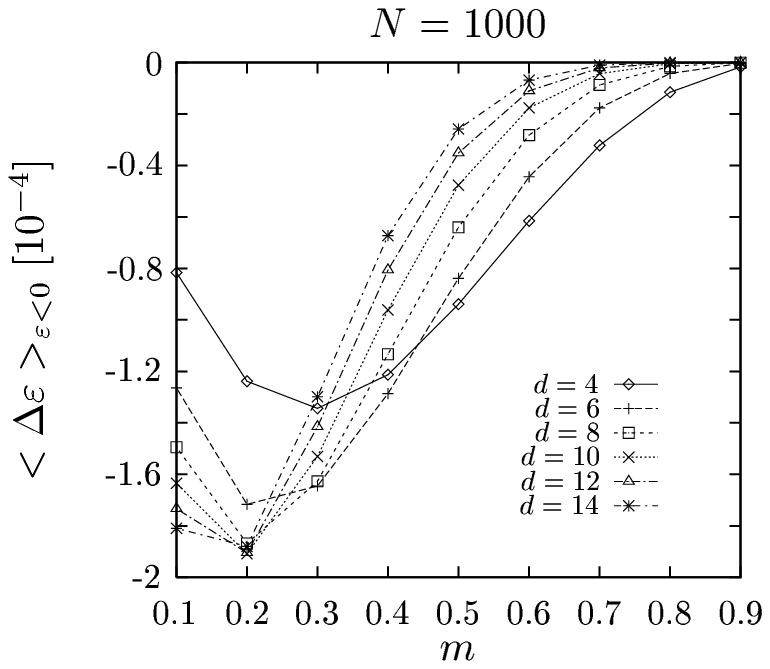, width=0.475\textwidth, height=5cm} 
\par}
\caption[]{Average energy change $\langle \Delta \varepsilon \rangle$ through
accepted moves in OMCD as a function of the magnetization $m$ for the 
Curie Weiss model, for various sizes $d$ of the move class. Left: system size
$N=100$. Right: system size $N=1000$. Evaluation according to Eqs (2) -- (6).}
\end{figure}

\begin{figure}[t]
{\centering
\epsfig{file=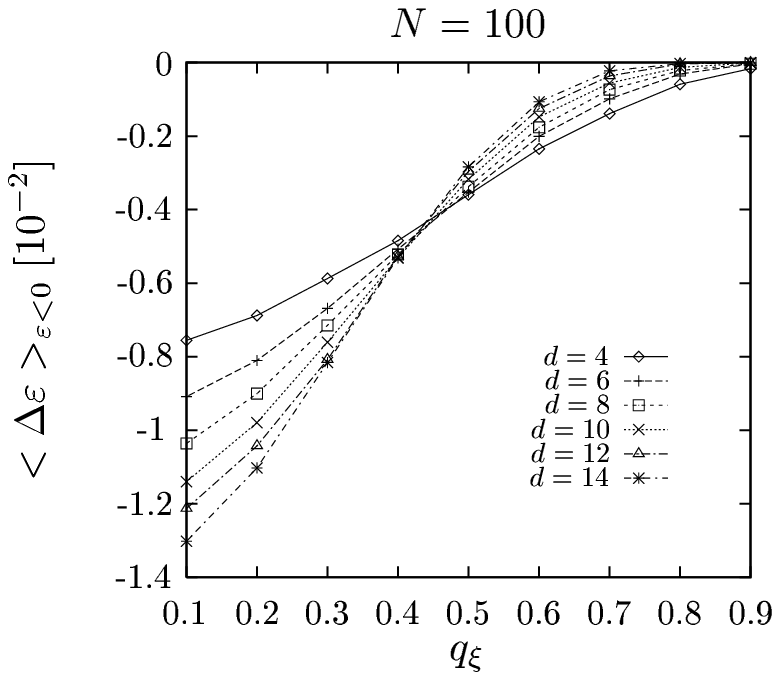, width=0.475\textwidth, height=5cm} 
\hfill{}
\epsfig{file=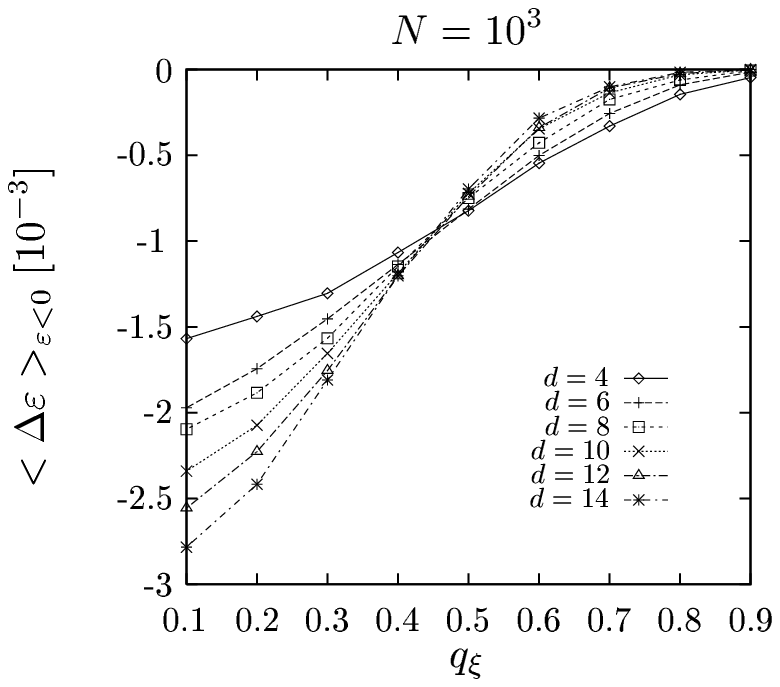, width=0.475\textwidth, height=5cm} 
\par}
\caption[]{Simulation results for average energy change $\langle \Delta \varepsilon \rangle$ through accepted moves in OMCD as a function of the overlap
$q_\xi$ with a ground--state configuration for the SK model (left) and the $\pm J$ model (right), for various sizes $d$ of the move class. The system sizes are
$N=100$ and $N=1000$, respectively.}
\end{figure}

Note that the ``effective" dimension of the accessible configuration
space goes down as the search  proceeds, and entropy/DOS arguments
might become less forceful as we are homing in on low energy configurations.
One might therefore consider the possibility of switching the strategy
towards an an exaustive search of the configuration space (within a small
Hamming distance of some state already reached) in the final stage of an
optimization run.

The above considerations can be verified quantitatively for the toy-problem
of finding the ground state of a Curie-Weiss ferromagnet
\be
\cH_N = -\Nh m^2\ ,
\ee
with $m$ denoting the magnetisation (per spin) of the system. For a move
class of size $d$, the probability to select $r$ out of $d$ spins with
$S_i=-1$ is given by
\be
\cP_r = {d \choose r} p^r (1-p)^{d-r}\ ,\quad {\rm with} 
\quad p=\frac{1-m}{2}\ .
\label{pr}
\ee
Flipping $r$ negative spins changes the magnetization (per spin) by
\be
\Delta m = \frac{2}{N} (2 r - d)
\ee
and thus the energy (per spin) by
\be
\Delta \varepsilon = -\Eh (\Delta m)^2 - m \Delta m \ .
\ee
We can express $\Delta m$ (hence $r$) in terms of $m$ and $\Delta \varepsilon$
\be
\Delta m = - m + \sqrt{m^2 - 2 \Delta \varepsilon}
\ee
which allows us, using (\ref{pr}), to compute the average decrease in energy $\langle \Delta \varepsilon \rangle$ from moves of size $d$ within the MCD 
algorithm as a function of the magnetization $m$. (To evaluate the average, 
we use the Stirling approximation $k!\simeq \sqrt{2\pi k} k^k \exp\{-k + 
\frac{1}{12 k}\}$ to compute factorials of large numbers.) The result is shown 
in Fig. 1, and it demonstrates that, depending on the distance to the 
minimum of the energy it appears to be advantageous to change from larger to 
smaller moves as the fully magnetized configuration (i.e. the state of minimal 
energy) is approached. In Fig. 2 we show for comparison that qualitatively 
similar results are obtained for the SK spin-glass and for the 3-$D$ $\pm J$ 
model, if the magnetization is replaced by the overlap $q_\xi = N^{-1} \sum_i 
\xi_i S_i$ between the system state $\vS$ and a ground-state configuration 
$\{\xi_i\}$.

\section{Scaling Properties}

In OMCD, the move--class deflation schedule plays a role analogous to the
annealing schedule in simulated annealing. Of primary interest here is
the  dize $d=d_0$ of the initial moves required to find good ground states
in the end. Clearly the initial size of the move-class should be chosen
such as to allow to jump over (or perhaps more appropriately tunnel through)
the widest energy barriers wich might typically separate an initially chosen 
random configuration from a good ground--state configuration.

\begin{figure}[h] 
{\centering \epsfig{file=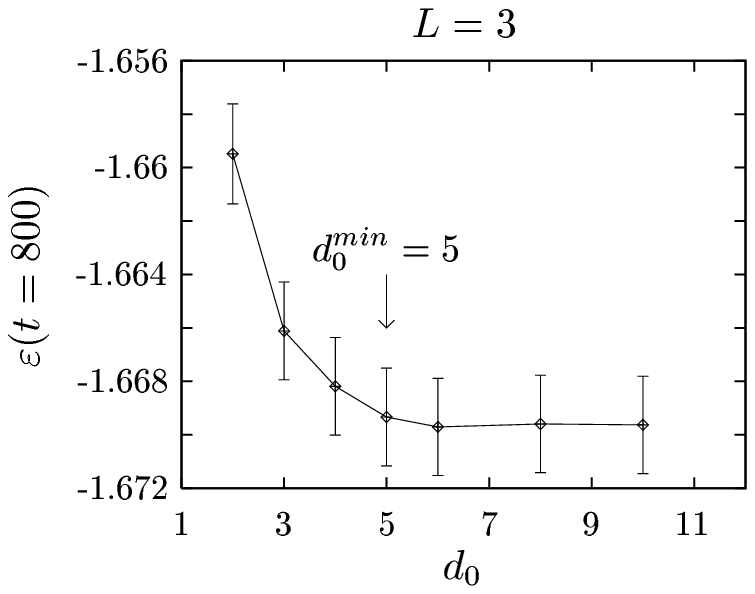, width=0.475\textwidth} 
\hfill 
\epsfig{file=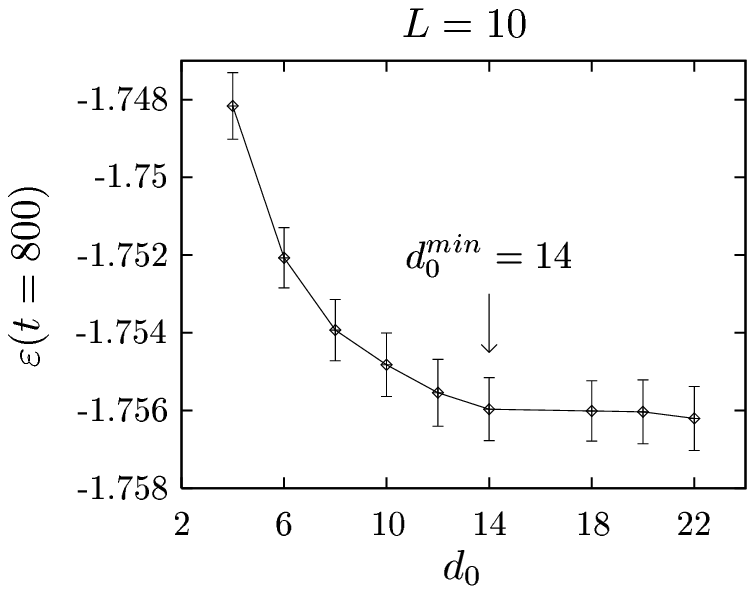, width=0.475\textwidth} \par}

\caption[]{Best energies, obtained with $t=800$ MCS at each $d$, for the $\pm J$
model of size $N=L^3$ as a function of the initial size $d_0$ of the move-class. Left: $L=3$, right: $L=10$. Results are averages over $M=4000$ samples for $L=3$, and over $M=350$ for $L=10$, and are displayed along with statistical 
errors $\sigma =\sqrt{{\rm var}(\varepsilon)/M}$.}
\end{figure}

\begin{figure}[h]
{\centering \epsfig{file=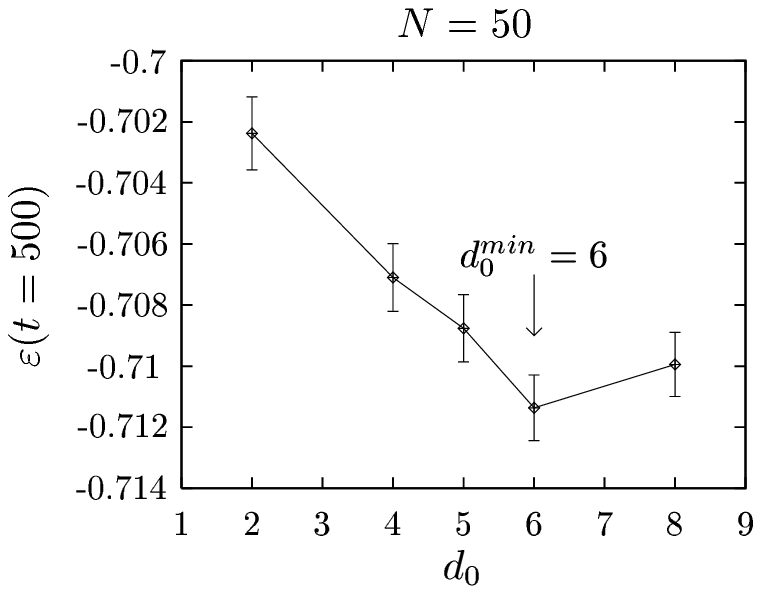, width=0.475\textwidth} 
\hfill 
\epsfig{file=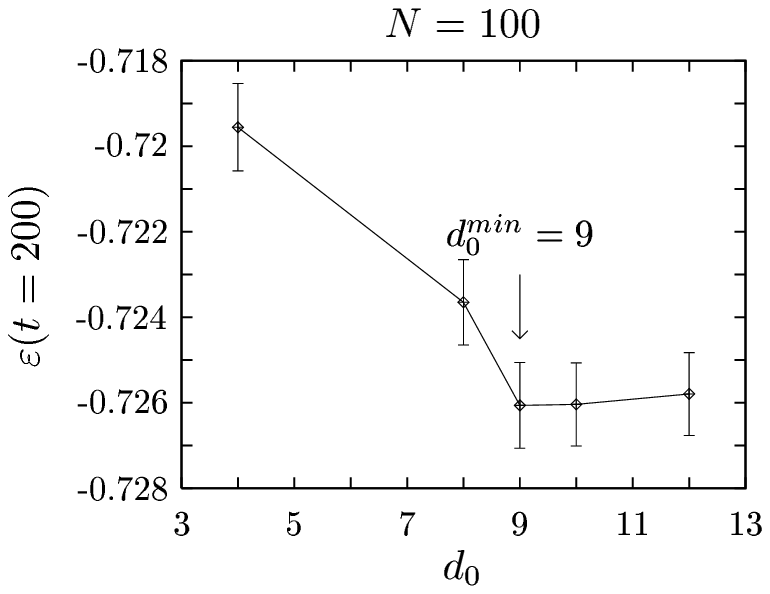, width=0.475\textwidth} \par}

\caption[]{Best energies $\varepsilon(t)$, obtained with $t$ MCS at each $d$, 
for the SK model of sizes $N=50$ (left) and $N=100$ (right) as a function of 
the initial size $d_0$ of the move-class. Results are displayed along with statistical errors. Averages are performed over 2000 and 500 samples
respectively.}
\end{figure}

In spin--glasses, the barrier--heights are well known to scale with system
size. For instance, in case of the SK model numerical \cite{VeVi} and 
analytical \cite{Ki,KiHo} investigations indicate a divergence of the barrier 
heights $\Delta E$ with system size $N$ according to 
\be
\Delta E \sim N^\lambda\ ,\quad {\rm with} \quad \lambda\simeq \frac{1}{3}\ .
\ee
We are currently not aware of quantitative studies of this issue for the
$\pm J$ model, although some divergence with $N$ is to be expected for this
case as well. If the energy landscapes of the SK and $\pm J$ models are 
`normal' in the sense described above, we have to anticipate a divergence of 
the barrier--widths with system size as well. Within OMCD this should manifest 
itself through the fact that we need a {\em minimal\/} initial size $d_0$
of the move--class, in order to arrive at good low--energy configurations
as the algorithm proceeds.

Figs. 3 and 4 show results for final energies obtained by OMCD as a function
of the initial size $d_0$ of the move class for $\pm J$ and SK models of
different sizes. From such simulations, we can extract $d_0^{min}$ as a function of system size $N$, and extract its scaling with $N$. For both cases we have
attempted two types of fit, using the {\tt mrqmin} routine from Press et al.
\cite{Pre+} (see Fig. 5), namely
\bea
d_0^{min} & = & a_1 N^{b_1} \\
d_0^{min} & = & a_2 + b_2 \ln N
\eea
for large $N$.

\begin{table}[h]
{\centering \begin{tabular}{crr}
\hline 
& \( \pm J \)-model&SK-model\\
\hline 
\hline 
 \( a_{1} \)& \( 2.79\pm 0.61 \)& \( 0.94\pm 0.55 \)\\
 \( b_{1} \)& \( 0.23\pm 0.04 \)& \( 0.49\pm 0.12 \)\\
 \( a_{2} \)& \( -1.90\pm 1.77 \)& \( -11.19\pm 4.55 \)\\
 \( b_{2} \)& \( 5.11\pm 0.77 \)& \( 10.12\pm 2.21 \)\\
\hline 
\end{tabular}\par}
\caption[]{Values for the fit parameters appearing in Eqs. (8) and (9).}
\end{table}

\begin{figure}[t]
{\centering 
\epsfig{file=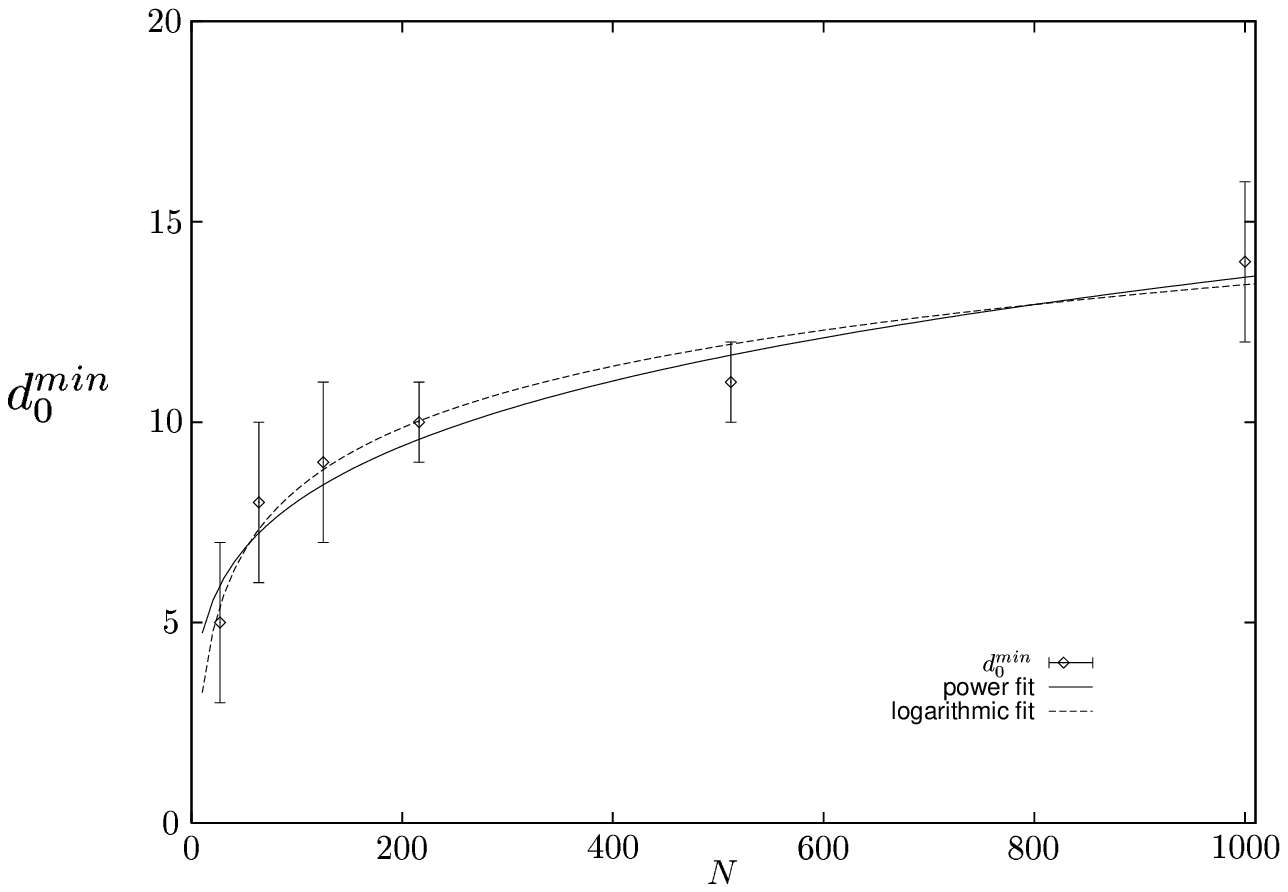, width=0.475\textwidth} 
\hfill
\epsfig{file=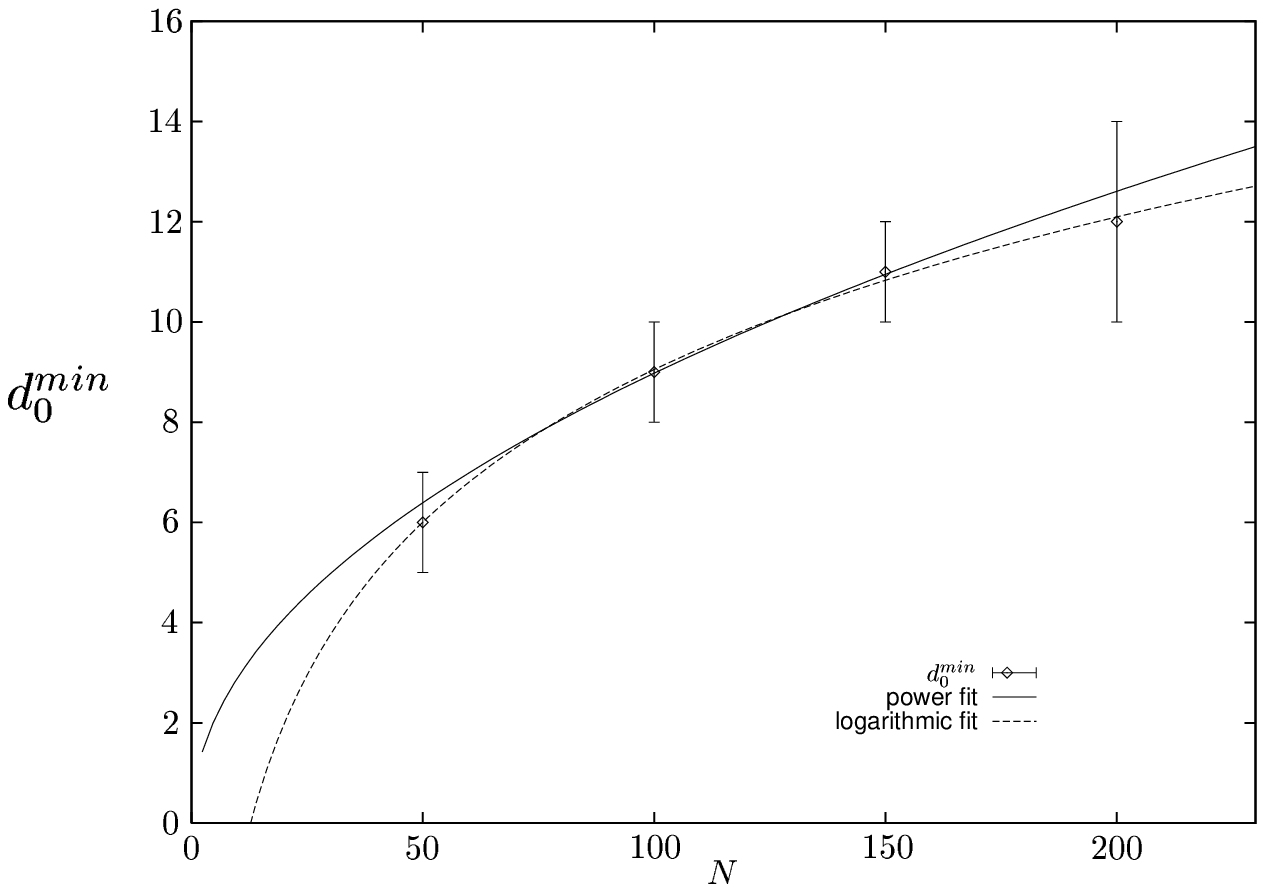, width=0.475\textwidth}
\par}
\caption[]{$d_0^{min}$ as a function of system size $N$ for the $\pm J$ 
model (left) and the SK model (right), with power-law and logarithmic fits.}
\end{figure}

The results are collected in Table 1 and Fig. 5. Somewhat to our 
surprise, we find that the logarithmic fits are consistently (though only 
slightly) better than the power--law fits. We obtained $Q_{\pm J}= 0.76$ and 
$Q_{SK}= 0.98$ for the logarithmic fits and $Q_{\pm J}= 0.65$ and $Q_{SK}= 
0.86$ for the power--law fit as measured by the {\tt mrqmin} routine 
\cite{Pre+}. It would -- at least for the SK model --- be difficult to 
reconcile this with a `normality' assumption about the 
energy landscape of this model, if we accept the fact that we have 
to tackle energy barriers scaling with system size $N$ as indicated in (7). 
However, the following observation might
be advanced in favour for the correctness of the logarithmic scaling anyway.
It is not excluded, in particular for the large system sizes, that OMCD
does not have to surmount the largest (hence widest) energy barriers at all, because due to the exponential scaling (in $N$) of the local DOS, the random 
initialization would with sufficiently high probability already select a 
state near the surface of the deepest and widest valley, so that the largest
and widest energy barriers need in fact not be passed through at all. In this 
context one might recall that Kinzelbach and Horner \cite{Ki,KiHo} observed 
that the spectrum of relaxation times for the finite SK model contains one 
with the weakest system-size dependence, scaling as $\tau \sim N^\nu$ with 
system size $N$. Assuming such a scaling to be due to an Arrhenius type 
mechanism with an activation energy barrier $B_N$, that is $\tau \sim 
\exp\{B_N/k_B T\}$, one would have to conclude that this type of barrier 
exhibits a logarithmic scaling with system size $B_N \sim \ln N$, and it is 
possible that we have to tackle only these very weakly diverging barriers in 
OMCD, as the system size becomes large. This point clearly deserves a deeper
and more extensive study.

\section{Ground--State Energies of Spin--Glasses}

We have used OMCD to determine ground state energies of the 3-$D$ $\pm J$
spin-glass and of the SK model. Following \cite{GreSou}, we have recorded the
lowest energies found by the MCD algorithm as a function of the time $t$,
measured by the number of attempted moves per spin (MCS) spent at each 
move-class size $d$. In analogy to the findings of these authors, we observe
a logarithmic dependence of the form
\be
\varepsilon(t) = \varepsilon_{min} + c_1 / \ln t\ ;
\ee
see Figs. 6 and 7. This is due to the fact that the problem of finding true 
ground--states for these systems is believed to be NP--hard.

\begin{figure}[h]
{\centering 
{\epsfig{file=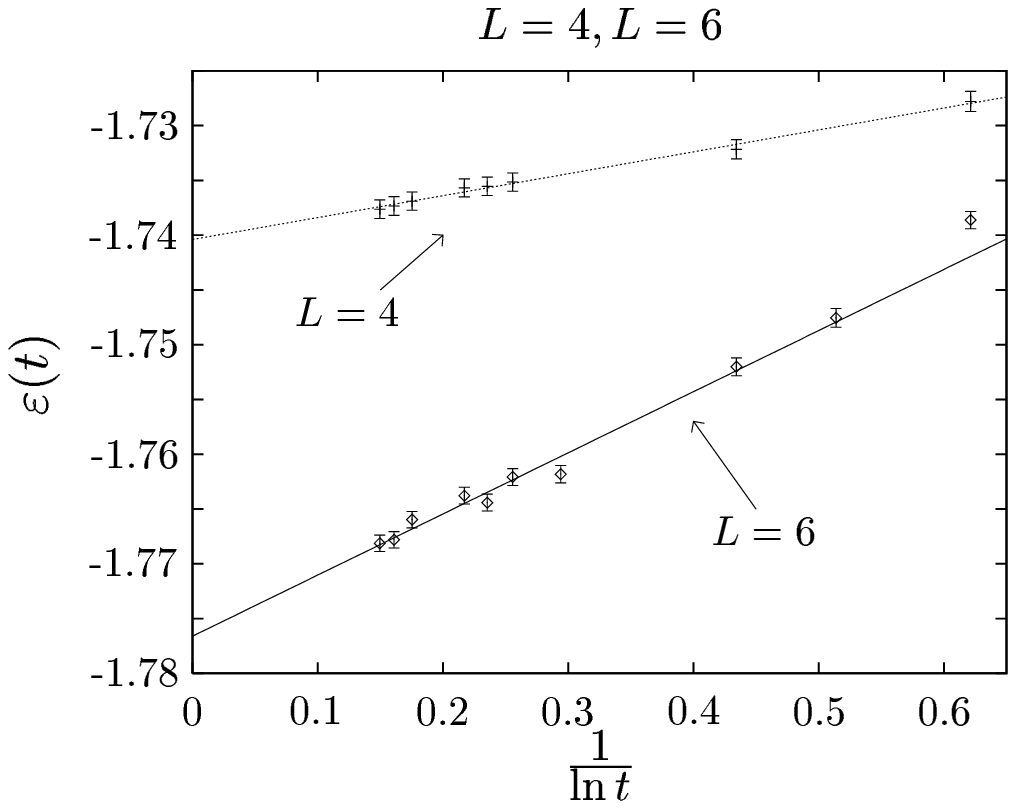, width=0.45\textwidth}
\hfill{}  
\epsfig{file=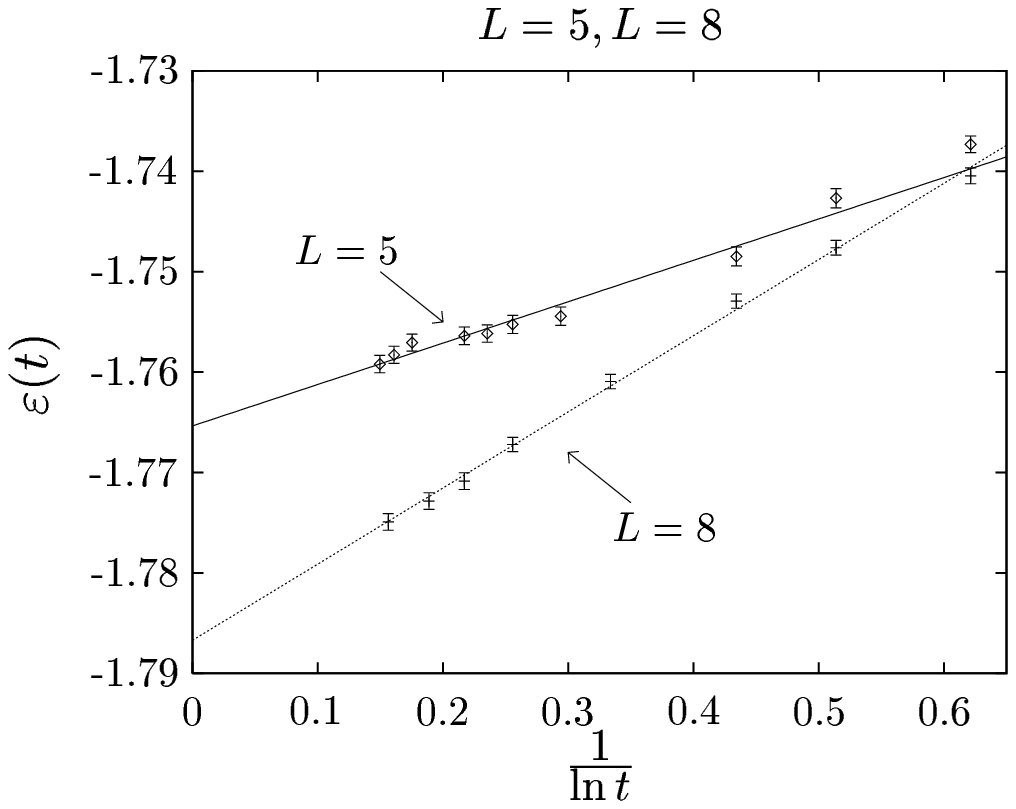, width=0.45\textwidth}}
\par}
\caption[]{Best energies obtained for $\pm J$ models of various sizes as a function of the time $t$ measured in MCS spent at each size $d$ of the 
move-class, exhibiting the logarithmic scaling of Eq. (10).}
\end{figure}

\begin{figure}h]
{\centering 
{\epsfig{file=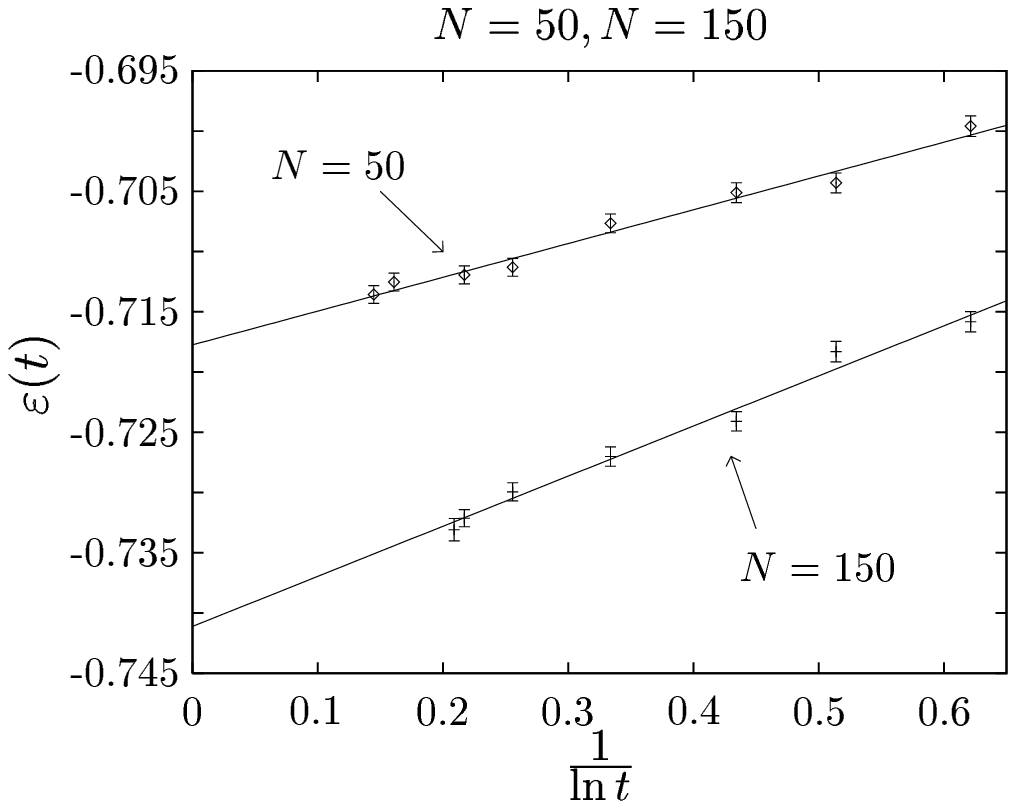, width=0.45\textwidth}
\hfill{}  
\epsfig{file=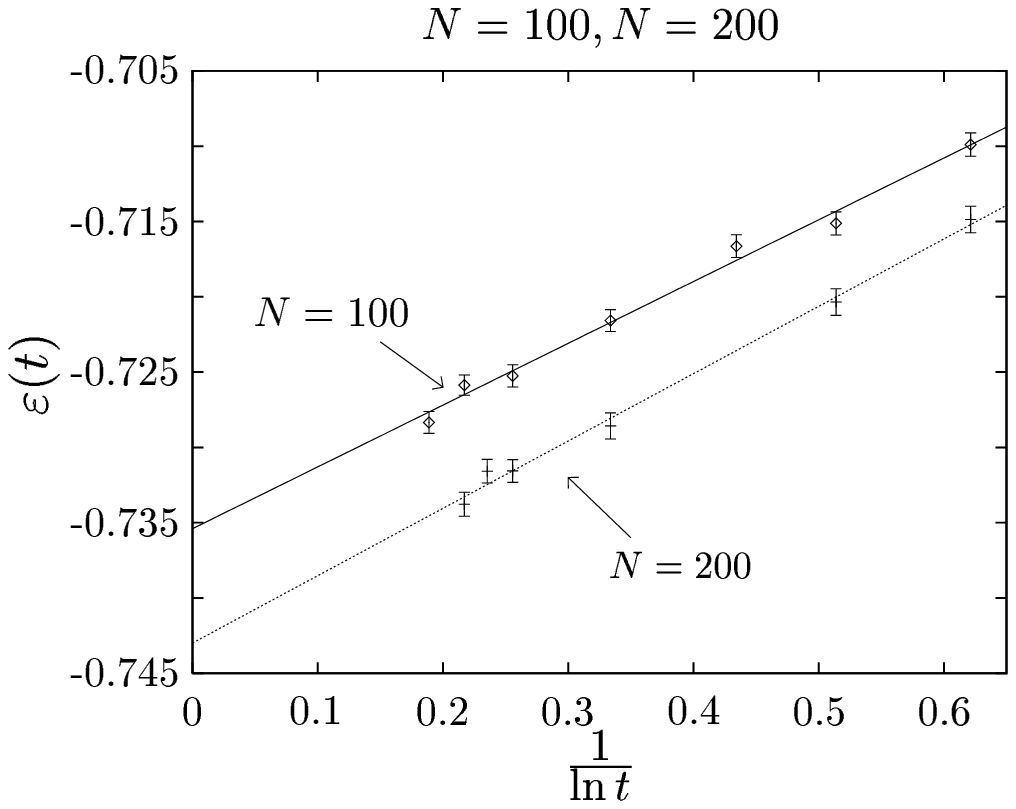, width=0.45\textwidth}}
\par}
\caption[]{Best energies obtained for SK models of various sizes as a function of the time $t$ measured in MCS spent at each size $d$ of the  move-class.}
\end{figure}

\begin{figure}[t]
{\centering 
\epsfig{file=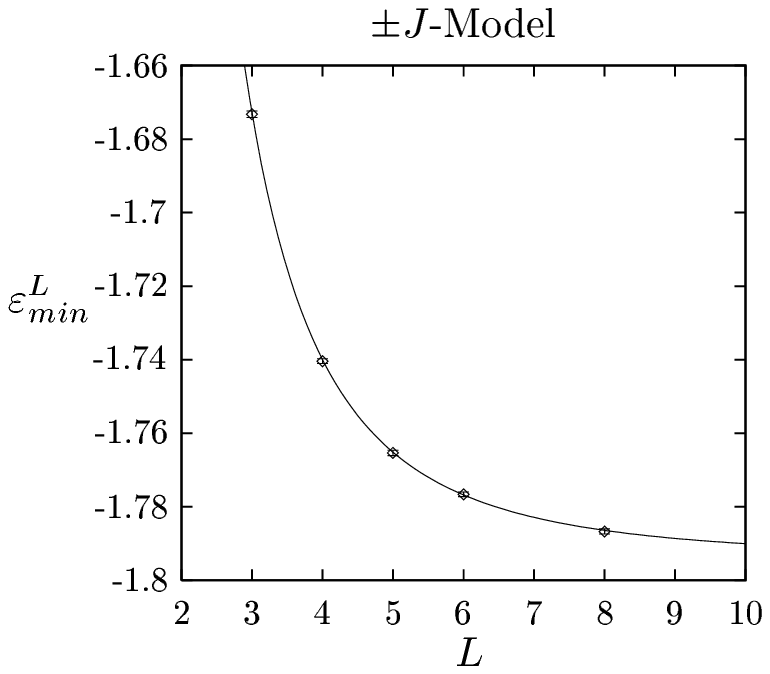, width=0.45\textwidth,height=6.5cm}\hfill{}
\epsfig{file=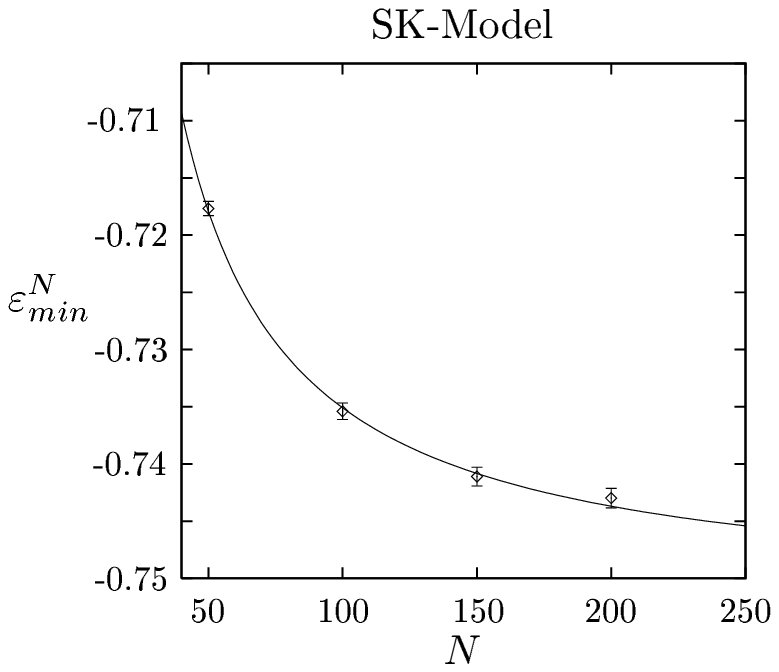,width=0.45\textwidth, height=6.5cm} 
\par}
\caption[]{Finite size scaling of the ground state energies of the $\pm J$ model
and the SK model, respectively.}
\end{figure}

\begin{table}[p]
{\centering \begin{tabular}{|c||c|c|c|c||c|c|c|c|}
\hline 
&\multicolumn{8}{|c|}{3D  \( \pm J \) -model}\\
\hline 
\hline 
&\multicolumn{4}{|c||}{  \( \qquad L=3 \) ( \( I_{N}=3,d_{0}=6 \))}&\multicolumn{4}{|c|}{  \( \qquad L=4 \) ( \( I_{N}=4,d_{0}=8 \))}\\
\hline 
 \( t \)& \( M_{N} \)& \( \varepsilon _{N}(t) \)& \( \sigma (\varepsilon ) \)& \( \tau  \)(s)& \( M_{N} \)& \( \varepsilon _{N}(t) \)& \( \sigma (\varepsilon ) \)& \( \tau  \)(s)\\
\hline 
\hline 
5&18000&-1.667647&0.000887&0.03&5000&-1.727787&0.000921&0.09\\
\hline 
10&18000&-1.670881&0.000863&0.06&5000&-1.732162&0.000880&0.18\\
\hline 
50&18000&-1.673235&0.000866&0.26&5000&-1.735146&0.000832&0.90\\
\hline 
70&---&---&---&---&5000&-1.735542&0.000841&1.25\\
\hline 
100&---&---&---&---&5000&-1.735687&0.000828&1.78\\
\hline 
300&18000&-1.673009&0.000826&1.65&5000&-1.736896&0.000839&5.32\\
\hline 
500&---&---&---&---&5000&-1.737333&0.000857&8.87\\
\hline 
800&---&---&---&---&5000&-1.737625&0.000849&14.18\\
\hline 
1000&18000&-1.672839&0.000865&5.50&---&---&---&---\\
\hline 
\hline 
&\multicolumn{4}{|c||}{
 \( \qquad L=5 \) ( \( I_{N}=4,d_{0}=10 \))}&\multicolumn{4}{|c|}{  \( \qquad L=6 \) ( \( I_{N}=5,d_{0}=12 \))}\\
\hline 
 \( t \)& \( M_{N} \)& \( \varepsilon _{N}(t) \)& \( \sigma (\varepsilon ) \)& \( \tau  \)(s)& \( M_{N} \)& \( \varepsilon _{N}(t) \)& \( \sigma (\varepsilon ) \)& \( \tau  \)(s)\\
\hline 
\hline 
5&3000&-1.737312&0.000836&0.39&2000&-1.738620&0.000781&0.86\\
\hline 
7&2000&-1.742681&0.000957&0.50&1500&-1.747556&0.000855&1.25\\
\hline 
10&2000&-1.748480&0.000945&0.71&1500&-1.752012&0.000819&1.71\\
\hline 
30&2000&-1.754432&0.000923&2.13&1500&-1.761815&0.000793&5.11\\
\hline 
50&2000&-1.755248&0.000907&3.60&1500&-1.762077&0.000769&8.49\\
\hline 
70&2000&-1.756168&0.000864&5.04&1500&-1.764412&0.000767&11.90\\
\hline 
100&2000&-1.756384&0.000876&7.12&1500&-1.763781&0.000768&17.02\\
\hline 
300&2000&-1.757060&0.000847&21.36&1500&-1.765975&0.000749&51.00\\
\hline 
500&2000&-1.758281&0.000854&35.58&1500&-1.767815&0.000741&85.00\\
\hline 
800&2000&-1.759207&0.000862&57.20&1500&-1.768124&0.000743&\multicolumn{1}{c|}{135.93}\\
\hline 
\hline 
&\multicolumn{4}{|c||}{
 \( \qquad L=8 \) ( \( I_{N}=15,d_{0}=12 \))}&\multicolumn{4}{|c|}{ }\\
\cline{1-1} \cline{2-5} 
 \( t \)& \( M_{N} \)& \( \varepsilon _{N}(t) \)& \( \sigma (\varepsilon ) \)& \( \tau  \)(s)&\multicolumn{4}{|c|}{ }\\
\cline{1-1} \cline{2-2} \cline{3-3} \cline{4-4} \cline{5-5} 
5&700&-1.740453&0.000784&1.73&\multicolumn{4}{|c|}{}\\
\cline{1-1} \cline{2-2} \cline{3-3} \cline{4-4} \cline{5-5} 
7&700&-1.747612&0.000735&2.41&\multicolumn{4}{|c|}{}\\
\cline{1-1} \cline{2-2} \cline{3-3} \cline{4-4} \cline{5-5} 
10&700&-1.752922&0.000710&3.45&\multicolumn{4}{|c|}{}\\
\cline{1-1} \cline{2-2} \cline{3-3} \cline{4-4} \cline{5-5} 
20&700&-1.760937&0.000716&6.92&\multicolumn{4}{|c|}{}\\
\cline{1-1} \cline{2-2} \cline{3-3} \cline{4-4} \cline{5-5} 
50&700&-1.767210&0.000702&17.48&\multicolumn{4}{|c|}{}\\
\cline{1-1} \cline{2-2} \cline{3-3} \cline{4-4} \cline{5-5} 
100&500&-1.770859
&0.000835&34.45&\multicolumn{4}{|c|}{}\\
\cline{1-1} \cline{2-2} \cline{3-3} \cline{4-4} \cline{5-5} 
200&500&-1.772844&0.000835&68.75&\multicolumn{4}{|c|}{}\\
\cline{1-1} \cline{2-2} \cline{3-3} \cline{4-4} \cline{5-5} 
600&500&-1.774922&0.000834&207.08&\multicolumn{4}{|c|}{}\\
\hline 
\end{tabular}\par}
\caption[]{Lowest energies of $\pm J$--models found, as a function of
$t$ measured in units of MCS/spin, and system size $N=L^3$. We also record
the statistical error and the average time needed for a {\em single\/} OMCD 
run on a 100 MHz 486DX-PC. $M_N$ is the numer of bond configurations used
for averaging, $I_N$ is the number of runs at each bond configuration, from
which the minimum was selected.}
\end{table}

\begin{table}[t]
{\centering \begin{tabular}{|c||c|c|c|c||c|c|c|c|}
\hline 
&\multicolumn{8}{|c|}{SK-model}\\
\hline 
\hline 
&\multicolumn{4}{|c||}{  \( \qquad N=50 \) ( \( I_{N}=5,d_{0}=6 \))}&\multicolumn{4}{|c|}{  \( \qquad N=100 \) ( \( I_{N}=8,d_{0}=9 \))}\\
\hline 
 \( t \)& \( M_{N} \)& \( \varepsilon _{N}(t) \)& \( \sigma (\varepsilon ) \)& \( \tau  \)(s)& \( M_{N} \)& \( \varepsilon _{N}(t) \)& \( \sigma (\varepsilon ) \)& \( \tau  \)(s)\\
\hline 
\hline 
5&2000&-0.699584&0.000857&0.22&1000&-0.709894&0.000775&1.77\\
\hline 
7&2000&-0.704299&0.000857&0.30&1000&-0.715125&0.000765&2.51\\
\hline 
10&2000&-0.705108&0.000824&0.43&1000&-0.716640&0.000756&3.53\\
\hline 
20&2000&-0.707653&0.000783&0.87&1000&-0.721576&0.000728&7.02\\
\hline 
50&2000&-0.711303&0.000748&2.15&1000&-0.725257&0.000740&17.77\\
\hline 
100&2000&-0.711930&0.000750&4.26&1000&-0.725867&0.000670&35.49\\
\hline 
200&---&---&---&---&800&-0.728344&0.000730&70.89\\
\hline 
500&2000&-0.712526&0.000745&21.00&---&---&---&---\\
\hline 
1000&2000&-0.713555&0.000733&42.40&---&---&---&---\\
\hline 
\hline 
&\multicolumn{4}{|c||}{
 \( \qquad N=150 \) ( \( I_{N}=15,d_{0}=11 \))}&\multicolumn{4}{|c|}{  \( \qquad N=200 \) ( \( I_{N}=20,d_{0}=12 \))}\\
\hline 
 \( t \)& \( M_{N} \)& \( \varepsilon _{N}(t) \)& \( \sigma (\varepsilon ) \)& \( \tau  \)(s)& \( M_{N} \)& \( \varepsilon _{N}(t) \)& \( \sigma (\varepsilon ) \)& \( \tau  \)(s)\\
\hline 
\hline 
5&500&-0.715820&0.000832&5.98&300&-0.714866&0.000872&12.35\\
\hline 
7&500&-0.718296&0.000849&8.23&300&-0.720348&0.000885&17.30\\
\hline 
10&500&-0.724086&0.000794&11.72&---&---&---&---\\
\hline 
20&500&-0.727007&0.000802&23.19&300&-0.728572&0.000866&48.87\\
\hline 
50&500&-0.729938&0.000758&58.12&300&-0.731563&0.000754&121.50\\
\hline 
70&---&---&---&---&300&-0.731578&0.000784&171.66\\
\hline 
100&500&-0.732130&0.000713&115.92&300&-0.733769&0.000795&243.89\\
\hline 
120&300&-0.733078&0.000928&139.12&---&---&---&---\\
\hline 
\end{tabular}\par}
\caption[]{Lowest energies of SK--models found, as a function of
$t$ measured in units of MCS/spin, and system size $N$. Symbols are
defined as in Table 2.}
\end{table}\

For the $\pm J$ model, we have investigated systems of size $L^3$, with 
$L= 3, 4, 5, 6$, and 8. For the SK model we have looked at system sizes 
$N= 50, 100, 150$, and 200. The initial size $d_0$ of the move--class was
chosen according to the results of the scaling-analysis of the previous
section. In the case of the $\pm J$ model, the $d$ spins to be flipped in a 
move-class of size $d$ were selected by a random walk which visited $d$ sites,
with starting point selected serially (or at random). In the case of the SK 
model, the $d$ spins were simply selected at random.

For each system (each bond configuration), $I_N$ optimization runs were 
performed, and only the absolute energy minimum among these runs was recorded.
Results were averaged over $M_N$ random bond configurations, with $M_N$ ranging 
from 18000 for the small systems to 500 for the large systems (and large $t$)
in the case of the $\pm J$ model, and similarly from from 2000 to 300 in the
case of the SK model. Our results are collected in Tables 2 and 3 (wich also
records the average time needed for a {\em single\/} OMCD run on a 100 MHz 
486DX-PC). 

\begin{table}[h]
{\centering \begin{tabular}{|c||c|c|c|c|c|c|}
\hline 
\multicolumn{7}{|c|}{3D \( \pm J \) -model}\\
\hline 
\hline 
 &\multicolumn{6}{|c|}{ \( \varepsilon ^{L}_{min} \) }\\
\hline 
 \( L \)& this work &Ref. \cite {GreSou}&Ref. \cite {Ber+} &Ref. \cite{Gro}&Ref.
\cite{Pal}&Ref. \cite{Har}\\
\hline 
3& \( -1.6732(9) \)&---&---& \( -1.68138 \)& \( -1.67171 \)& \( -1.6731 \)\\
\hline 
4& \( -1.7404(6) \)& \( -1.791 \)& \( -1.7378 \)& \( -1.73973 \)& \( -1.73749 \)& \( -1.7370 \)\\
\hline 
5& \( -1.7654(7) \)&---&---& \( -1.76101 \)& \( -1.76090 \)& \( -1.7603 \)\\
\hline 
6& \( -1.7766(6) \)&---& \( -1.7674 \)& \( -1.77059 \)& \( -1.77130 \)& \( -1.7723 \)\\
\hline 
7&---&---&---& \( -1.77842 \)& \( -1.77706 \)&---\\
\hline 
8& \( -1.7867(7) \)&---& \( -1.7799 \)& \( -1.77901 \)& \( -1.77991 \)& \( -1.7802 \)\\
\hline 
10&---&---&---&---& \( -1.78339 \)& \( -1.7840 \)\\
\hline 
12&---&---& \( -1.7936 \)&---& \( -1.78407 \)& \( -1.7851 \)\\
\hline 
14&---&---&---&---& \( -1.78653 \)& \( -1.7865 \)\\
\hline 
\end{tabular}\par}
\caption[]{Ground state energies for the 3--$D$ $\pm J$ model of various
sizes.}
\end{table}

\begin{table}[h]
{\centering \begin{tabular}{|c||c|c|}
\hline 
&\multicolumn{2}{|c|}{SK-model }\\
\hline 
\hline 
&\multicolumn{2}{|c|}{ \( \varepsilon ^{N}_{min} \) }\\
\hline 
 \( N \)&this work&Ref. \cite {GreSou}\\
\hline 
50& \( -0.7177(6) \)& \( -0.7174 \)\\
\hline 
100& \( -0.7354(7) \)& \( -0.7354 \)\\
\hline 
150& \( -0.7411(8) \)&---\\
\hline 
200& \( -0.7430(9) \)& \( -0.7472 \)\\
\hline 
400&---& \( -0.7534 \)\\
\hline 
800&---& \( -0.7591 \)\\
\hline 
1300&---& \( -0.7620 \)\\
\hline 
\end{tabular}\par}
\caption[]{Ground state energies for SK models of various sizes.}
\end{table}

Tables 4 and 5 contain $t\to\infty$--extrapolations of the lowest energies
found according to Eq. (10), and compare with results of Refs. \cite{GreSou}, 
\cite{Ber+} -- \cite{Har}. Note that $t\to \infty$ extrapolations were not
performed in \cite{Ber+} -- \cite{Har}. The results of Berg et al. \cite{Ber+} 
were obtained using multicanonical sampling, those of \cite{Gro,Pal,Har} by combining the genetic algorithm with some other strategy, whereas \cite{GreSou} 
used simulated annealing.

The finite-size signature of the true average ground state energies
was extracted to follow the scaling
\be
\varepsilon_{min}^L = \varepsilon_{min}^\infty + a_1 L^{-a_2}
\ee
with
\be
\varepsilon_{min}^\infty = -1.7942(10) \qquad , \qquad a_1 = 2.63(19) \qquad , \qquad a_2 = 2.80(7)
\ee
for the 3--$D$ $\pm J$ model on cubes of side--length $L$, and the scaling 
\be
\varepsilon_{min}^N = \varepsilon_{min}^\infty + \frac{b}{N}
\ee
with
\be
\varepsilon_{min}^\infty = -0.7523(8) \qquad , \qquad b = 1.72(6) 
\ee
for the SK model (the deviation of $1.5\%$ from the analytic result \cite{Pa80}
might be due to the fact that the groundstate energy of the largest stystem 
used in our fit is perhaps somewhat poor). Results along with the above fits
are shown in Fig. 8.

So far we have paid little attention to details of the deflation schedule
apart from its starting with a sufficiently large move--class, i.e. to the
{\em optimization of the optimization algorithm itself}. The above results were 
obtained using a linear decrease $d \to d-1$ of the size of the move-class,
staring from $d_0$ as determined in the previous section. Other deflation
schedules may clearly be considered. Here we briefly mention the `exponential'
schedule
\be
d \to [\gamma d]\ ,\quad {\rm with} \quad \gamma < 1\ ,
\ee
where $[x]$ denote the largest integer less than or equal to $x$. This schedule
is approximately exponential for large $d$,   and it crosses over to linear, as 
soon as $\gamma d > d-1$. We have compared the results of such a schedule with
$\gamma = 0.8$ on the $\pm J$ model of size $N=5^3$  and on the  SK model with 
$N=100$ for various $t$. The change in the estimate of the lowest energy is 
in the 0.2\% range, while the computational cost was measured to decrease by 
roughly 40\% in the case of the $\pm J$ model, and by roughly 30\% for the 
SK model! This little study shows that there may still be room for considerable 
improvements of the algorithm itself, improvements which might clearly benefit 
from tayloring in problem specific ways.

\section{Phase--Space Diagnostics}

As we have mentioned in our introduction, and also when describing the 
heuristics of OMCD, the MCD algorithm is sensitive to properties of phase
spaces of complex systems other than those seen by simulated annealing.
For instance, deep down in a narrow valley large moves will not be accepted
because they could only lead us uphill (or perhaps outside the valley we are
currently in). This observation might in the future be used to develop 
strategies for self-optimization of the move-class deflation schedule as
it proceeds  with any given optimization task.

Here we illustrate this feature by monitoring --- for the $\pm J$ and the
SK model (and also for a simple 3-$D$ ferromagnet ---- the number of accepted
moves per MCS which decrase the energy, as a function of the size $d$ of the 
move--class, which is deflated linearly. For the $\pm J$ model and the ferromagnet, we have seperately monitored the moves which are accepted but lead
to a degenerate state. Such moves with $\Delta E = 0$ in the $\pm J$ model and 
the ferromagnet occur predominantly at very small $d$, as seen in Fig 9.

\begin{figure}[t]
{\centering 
\epsfig{file=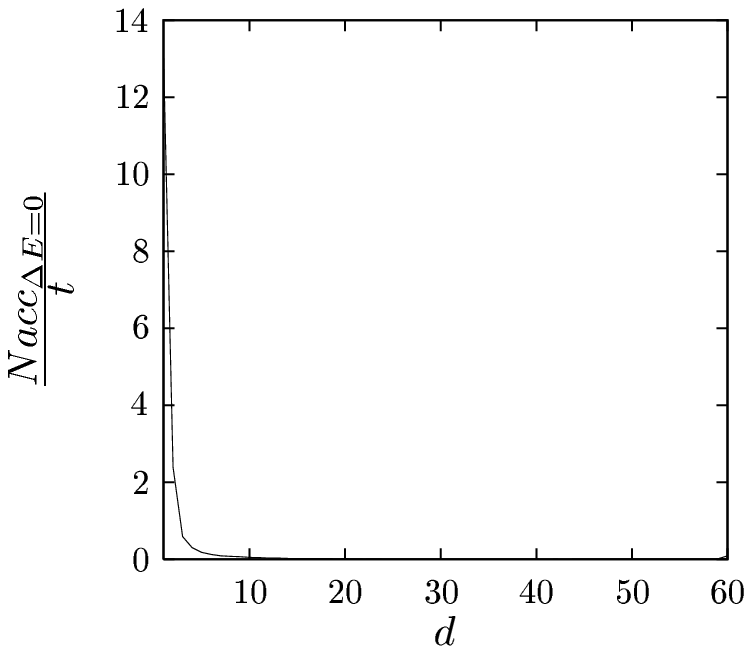, width=0.475\textwidth,height=5cm}\hfill{}
\epsfig{file=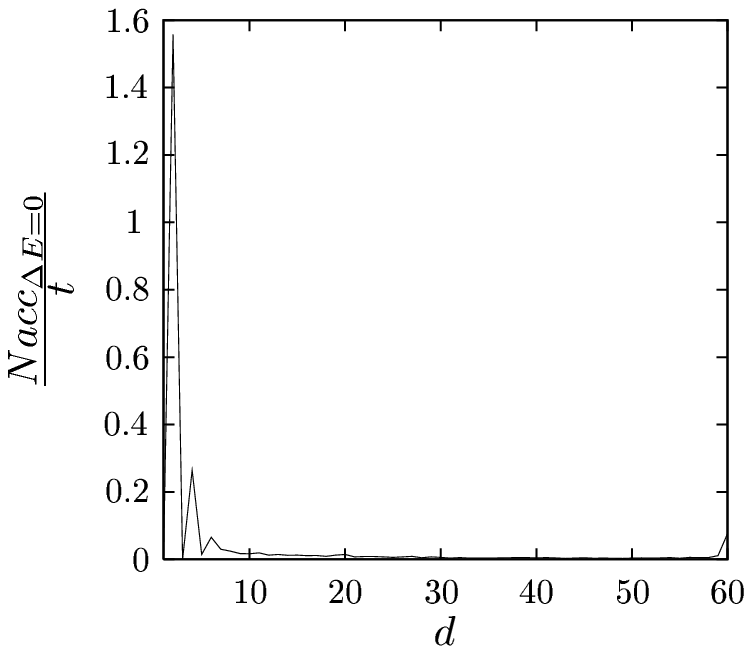,width=0.475\textwidth, height=5cm} 
\par}
\caption[]{Number of moves accepted per MCS with $\Delta E = 0$ in the $\pm J$ 
model (left) and the ferromagnet (right), as a function of the size $d$ of the 
move--class. The system size is $N=216$.}
\end{figure}

Fig. 10 shows clear differences between the SK model on the one hand side and
the $\pm J$ model and the ferromagnet on the other side. Apart from the 
`transient' behaviour near the initially chosen largest move class size, there
is a clear trend for the SK model towards acceptance of predominantly small
moves, whereas this feature is much less pronounced in the other two models.
On the other hand, there seems to be some systematic depression of the 
acceptance rate in the $\pm J$ model near $d=47$ and $d=40$, which survives 
the averaging over different bond-configurations implied in the representation 
of Fig. 10.

\begin{figure}[p]

{\centering \epsfig{file=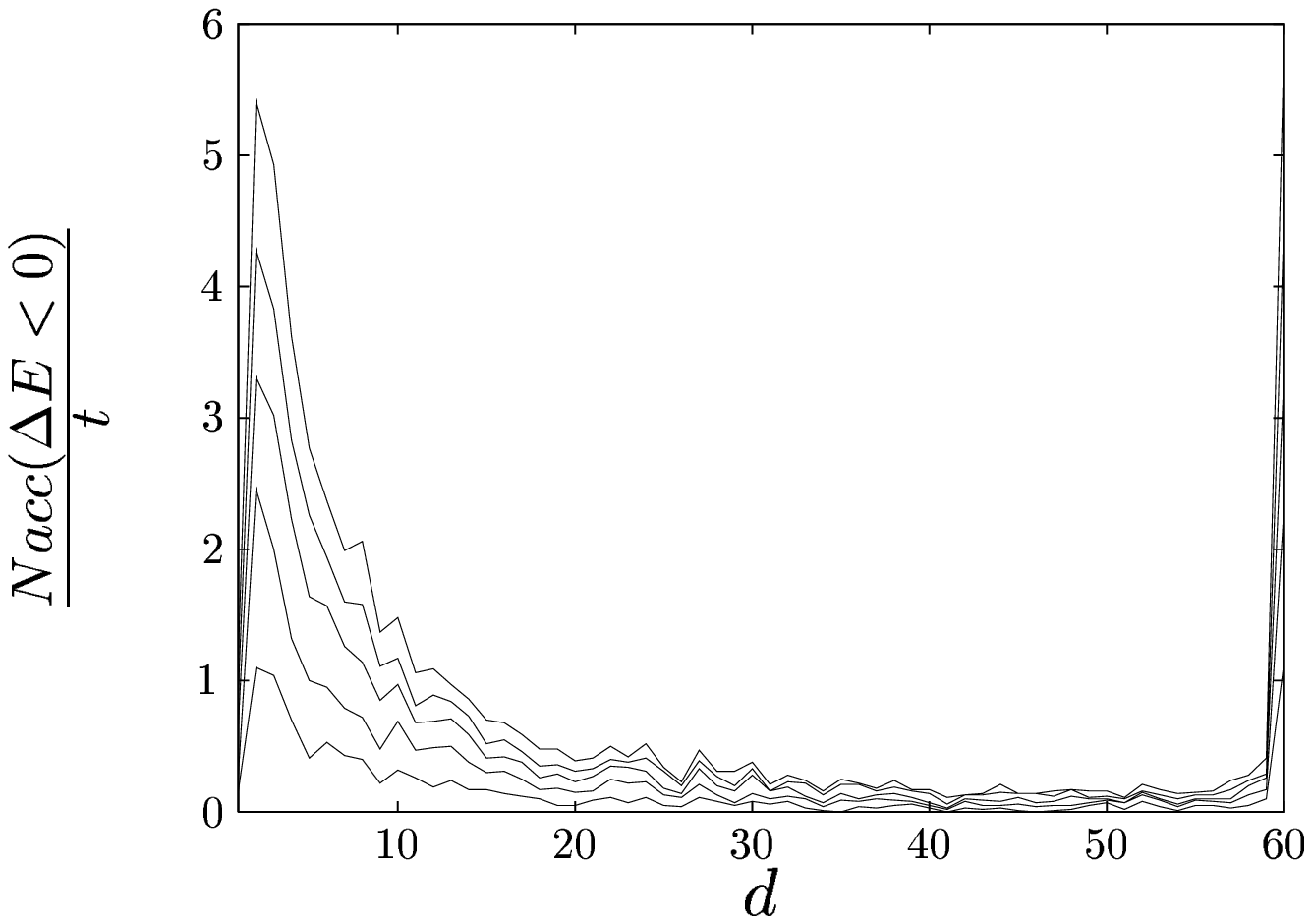, height=0.3\textheight} \par}
\vspace{0.5cm}
{\centering \epsfig{file=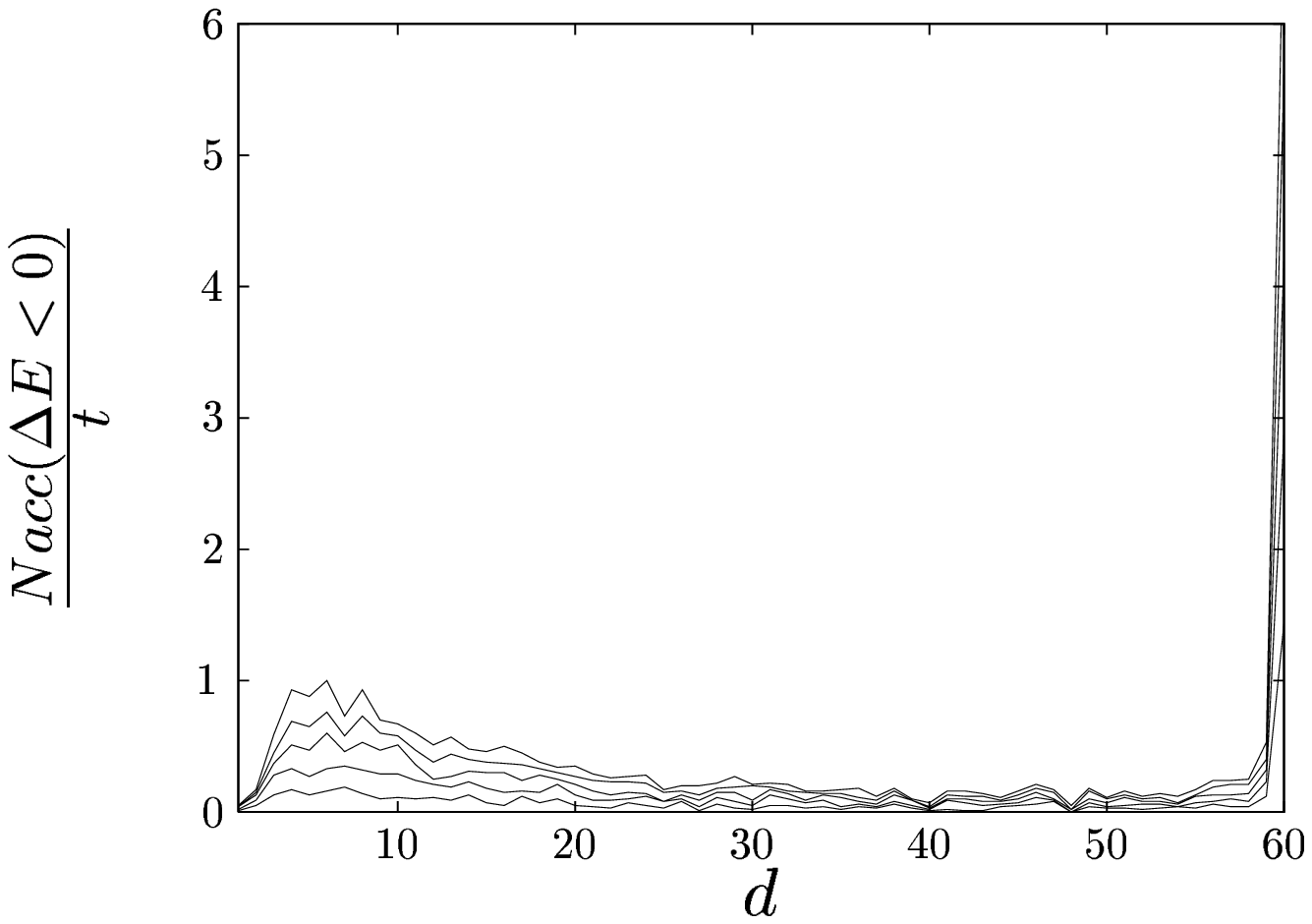, height=0.3\textheight} \par}
\vspace{0.5cm}
{\centering \epsfig{file=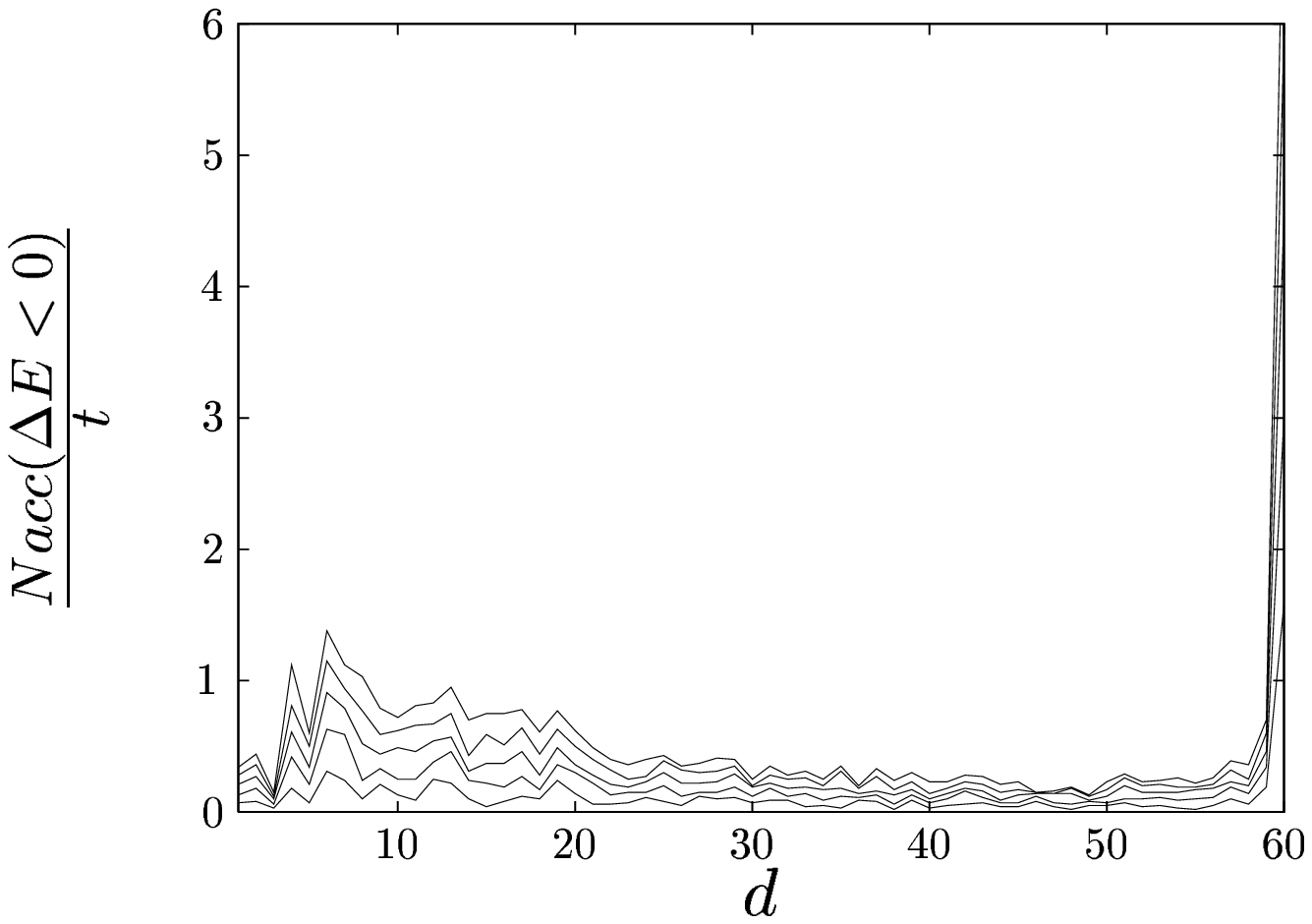, height=0.3\textheight} \par}

\caption[]{Number of moves accepted per MCS with $\Delta E < 0$ in the SK model,
the $\pm J$ model, and the ferromagnet (from top to bottom), as a function of the size $d$ of the move--class, accumulated over 10, 20, 30, 40, and 50 
samples. The system size is $N=216$ in all cases, and $t=100$.}
\end{figure}

Whether the difference between the SK model on the one hand side and the $\pm J$ model and the ferromagnet on the other side is due to the dicreteness of the
energy spectrum in the latter two models which is not shared by the SK model,
or due to the fact that the phase spaces of these models are in different 
complexity classes, we can at present not tell: It is well kown that the SK
model has a non-trivial distribution of overlaps between its various ground
states. For the $\pm J$ model this issue is currently still under debate, 
while in the ferromagnet, the overlap distribution is known to be trivial 
(i.e., to consist of two delta functions at $\pm 1$).

\section{Summary and Discussion}

In summary, we have proposed an alternative approach to complex combinatorial
optimization problems named OMCD, based on a systematic deflation of 
move--classes from mesoscopic to microscopic. The algorithm combines heuristics of genetic algorithms and simulated annealing. We expect it to
be efficient for problems with energy surfaces which are normal in the sense
that deep valleys are also wide valleys (and correspondingly high barriers
are also wide barriers). We have argued that it makes {\em constructive\/} 
use of the high dimensionality of phase spaces typically
encountered in large combinatorial optimization tasks, specifically of the
exponential scaling of (local) densities of state with system size $N$. That 
is, the algorithm {\em exploits\/} a feature which is generally believed to 
constitute a major difficulty to be tackled in combinatorial optimization.
It is perhaps worth mentioning that local densities of state do play a role 
also in the dynamics of other combinatorial optimization algorithms, if perhaps 
less explicitly (or less consciously) so.

We have verified our heuristics analytically on the toy--problem of finding
the ground state of a Curie--Weiss ferromagnet and numerically on the search
for ground-states in the three--dimensional $\pm J$ spin--glass and in the
SK model --- both problems believed to be NP-hard.

We estimate the time--complexity of the algorithm to be $\cO(tN (\log N)^2)$
for the $\pm J$ model and $\cO(tN^2(\log N)^2)$ the SK model, for the linear
move--class deflation schedule. We assume here correctness of the scaling
$d_0 =\cO(\log N)$ for the initial size $d_0$ of the move--class with system
size $N$. In case of the exponential move--class deflation schedule, the 
$(\log N)^2$ factors in the above estimates can be replaced by $\log N$ 
factors. 

We have seen that the exponential MCD schedule gives a significant reduction
of the computational cost without degrading the results. Clearly, further
improvements of these schedules are conceivable. Let us mention the possibility
of reducing the move--class size probabilistically, e.g. to have  distributions
of the form $\cP_{\overline d}(d) \propto \exp(-d/\overline d)$ for picking
a move--size $d$, and gradually reducing $\overline d$. This always gives
some scatter in the size of the move--class used at any time, which we have
observed to be advantageous. Different forms for $\cP_{\overline d}(d)$ may be
contemplated and might in some cases be expedient, which in particular exhibit 
{\em lower\/} cutoffs $d_{min} ({\overline d})$ --- gradually reduced to 1 
during move--class deflation --- so as to prevent the occurrence of small 
moves at the beginning of an OMCD run.

Moreover, the basic heuristics of the algorithm expects large moves to
be accepted in sufficiently wide valleys, and deeper down or in narrower energy valleys only smaller moves. This observation clearly lends itself to formulating
strategies for an algorithmic self--control of the move--class reduction schedule depending on recent acceptance rates, which might eventually lead to 
further improvements of OMCDs efficiency.

One of the advantages of OMCD over simulated annealing is in the
fact that it is not hampered by the freezing transitions in the same way as
the single spin flip Metropolis algorithms are, when applied to problems for
which the energy landscape is complex. Simulated annealing in its standard
implementations is therefore inefficient when used significantly below 
freezing temperatures in problems with glassy dynamics. So, to obtain good
low--lying energy configurations with such an algorithm, one is really 
`living on fluctuations'. The absence of freezing transitions in OMCD may
well become one of its decisive assets when dealing with large scale 
problems.

As OMCD is driven by properties of complex phase spaces in ways different
from simulated annealing or genetic algorithms, it may be used for phase space
diagnostics as demonstrated in Sect. 5. We have seen that the SK model appears
differently to MCD than the (presumably) simpler $\pm J$ model and the
ferromagnet. It should be noted however that we are only just beginning to
learn how to decipher the kind of analysis presented there.

\bigskip\noindent
{\bf Acknowlegdgement} We are indebted to A. Hartmann for helpful discussions,
and for providing us with some ground-state configurations for the $\pm J$ model
which allowed us to perform further tests of the quality of our results.

\end{document}